\documentclass[cameraready]{Interspeech}
\title{Rethinking Depth: A study of the Recursive-Transformer for Speech Recognition}

\author[affiliation={1}, orcid=0000-0003-2419-1377]{Thomas}{Rolland}
\author[affiliation={1,2}, orcid=0000-0002-9603-2423]{Carlos}{Carvalho}
\author[affiliation={1,2}, orcid=0000-0003-2122-5148]{Alberto}{Abad}
\address{
    $^1$ INESC-ID, Portugal \\
    $^2$ Instituto Superior Técnico, Portugal
}

\email{thomas.rolland@inesc-id.pt, carlos.carvalho@inesc-id.pt, alberto.abad@inesc-id.pt}

\keywords{Parameter-efficient, Latent-Recursive-Transformer, Recursive-Transformer, ASR, Layer-Sharing}

\usepackage{comment}

\begin{document}

\maketitle

\begin{abstract}
Transformer-based architectures have led to significant improvements in Automatic Speech Recognition (ASR), often at the cost of substantially increased model sizes. A promising approach to address this issue is layer sharing through depth recursion, commonly referred to as the Recursive-Transformer, which involves repeatedly applying the same layers within the model. Despite its potential shown in other fields, this technique remains relatively unexplored in ASR. In this paper, we present an experimental study of the Recursive-Transformer applied to ASR encoder architectures. We systematically investigate the impact of recursion
depth and layer allocation within the Recursive-based Transformer. Our results demonstrate that the Recursive-Transformer is a viable alternative, especially when recurrence is applied in the latent space with a restricted number of loops, obtaining comparable performance while reducing the parameter count by 66\%. 
\end{abstract}

\section{Introduction}
\label{sec:intro}
Recent Automatic Speech Recognition (ASR) advances are driven by scaling model size and datasets, with state-of-the-art systems now containing configurations that exceed one billion parameters and can be trained on millions of hours of speech data \cite{koluguri2025granary,radford2023robust, chen2025owls}.
As both model size and training data have continued to scale, performance improvements have followed accordingly. To formalise these empirical observations, researchers have introduced scaling laws \cite{Kaplan2020ScalingLF,chen2025owls}. These laws suggest that the performance of Transformer-based models \cite{Transformer} at scale can be predicted on the basis of three variables: model size, data size, and computational budget. Importantly, the computational budget is often directly tied to the number of model parameters, as larger models require proportionally more computation.

While increasing these three variables typically leads to enhanced performance, constraining any of these dimensions can produce the opposite effect. This limitation poses a significant challenge for on-device applications, where models are restricted by strict memory and storage constraints \cite{gondi2021performance}. Additionally, several studies have identified model size, or the number of parameters, as the primary bottleneck, often more restrictive than the amount of data or computational budget \cite{gholami2024ai,choi2024impact}.

To mitigate these challenges, model compression techniques have been developed \cite{cheng2017survey}, including knowledge distillation \cite{novitasari25b_interspeech}, pruning \cite{someki25_interspeech}, and quantization \cite{li25_interspeech}. Typically, model compression involves transferring knowledge from a large pre-trained model to a more compact version. However, two notable limitations remain. First, such techniques often rely on the existence of a large pre-trained model. Second, reducing the number of parameters inherently reduces the computational budget. While this reduction can be advantageous for time-constrained tasks, it simultaneously diminishes two of the three factors in scaling laws. This trade-off is particularly undesirable in high-accuracy regimes where maximizing recognition performance is more critical than minimizing inference latency \cite{alizadeh2024llm}.

An alternative direction is offered by parameter-sharing methods, which aim to reduce model size while maintaining computational budget, without the need for a large pre-trained model \cite{pham2018efficient, lan2019albert}. These approaches involve reusing parameters across multiple components of the architecture. This sharing may be applied at different granularities: weights of specific sub-modules across all layers of the Transformer \cite{pires2023one,rolland24b_interspeech}, entire layers individually \cite{UT2019,10.1007/978-3-031-20053-3_42, chi2021audio}, or groups of layers collectively \cite{wei2023sim, wang2024residualtransformer,yu2026spiralformer}, thereby introducing a recursive looping mechanism over the shared layers or groups.

Although these strategies are variously termed Universal-Transformers \cite{UT2019}, Recursive-Transformers \cite{bae2024relaxed,xu2026looping}, or Looped-Transformers \cite{fan2025looped} within the Natural Language Processing (NLP) literature, this study adopts the term Recursive-Transformer for consistency.
Recursive-Transformers have recently gained more attention within the  NLP field \cite{pires2023one,xue2022go,yu2026spiralformer}, particularly in the context of decoder-only large language models (LLMs). In contrast, their application to ASR remains limited \cite{wei2023sim,wang2024residualtransformer, shim24_interspeech,tang2024beyond, chi2021audio}, especially with regard to the encoder \cite{rolland2025exploring}. To address this gap, we investigate both the standard Recursive-Transformer \cite{UT2019} and a recent variant, the Latent-Recursive-Transformer \cite{geiping2025scaling}, in the context of ASR encoders. 
Particularly, this work systematically investigates the impact of recursion depth and layer allocation within the Recursive-based Transformer to identify the optimal balance between parameter efficiency and recognition accuracy. Our primary contributions are as follows:
\begin{itemize}
    \item We provide the first reported layer-similarity analysis of a large-scale ASR encoder, identifying redundant middle layers that justify the use of recurrence.
    \item We introduce the Latent-Recursive-Transformer for ASR, which uses a modular Prelude-Recurrent-Coda structure \cite{geiping2025scaling}.
    \item We demonstrate a 66\% parameter reduction while maintaining comparable performance to standard Transformers, even surpassing them in parameter-matched settings.
    \item We identify a critical trade-off in recurrence depth and evaluate how the number of loops interacts with various data characteristics.
    \item We confirm the robustness of the Latent-Recursive approach across diverse languages and architecture variants.
\end{itemize}

%Our analysis evaluates several key aspects of their design and performance.

\section{Related work}
\label{sec:related}
%Weight Sharing and Recurrence in Transformer Architectures
The Universal Transformer \cite{UT2019} was the first work to introduce recurrence in Transformer-based architectures \cite{Transformer} by iteratively applying a single layer to refine representations, effectively decoupling computational depth from the number of parameters. While recurrence is an established concept, it has seen a resurgence in recent literature following discoveries that these architectures exhibit enhanced reasoning capabilities in LLMs \cite{geiping2025scaling}. Specifically, Recursive-Transformer architectures have demonstrated superior performance in complex tasks by facilitating direct reasoning within the latent space \cite{zhu2025scaling,saunshi2025reasoning}, primarily through iterative depth-wise computations at the intermediate layers. This shift in perspective is supported by empirical analyses of representation spaces, which identify three distinct representation spaces: ``initial," ``middle," and ``final". In particular, the middle layers are largely interchangeable and permutable \cite{sun2024transformer, reid2021subformer}.

To bridge the expressivity gap in recursive-Transformer models, research has introduced layer-wise non-shared residual low-rank matrices \cite{wang2024residualtransformer,bae2024relaxed, tang2024beyond} or non-shared components \cite{shim24_interspeech}. Although recursive architectures drastically reduce the parameter footprint, they do not inherently mitigate inference latency. This limitation is addressed through Adaptive Early Stopping, early-exit or routing mechanisms \cite{graves2016adaptive,bae2025mixture, UT2019}, which allow the model to dynamically truncate recursion once a representation reaches a stability threshold.

Turning to speech, the ASR task represents a highly relevant application for recursive models, driven by the demand for on-device efficiency. Despite this potential, only a limited number of works have specifically targeted this task \cite{wei2023sim, wang2024residualtransformer, shim24_interspeech}. Notably, the Shared-Conformer and Shared-Transformer for ASR encoders \cite{rolland2025exploring} demonstrate that a single shared encoder layer can achieve substantial parameter reduction while remaining competitive in WER.

\section{Latent-Recursive-Transformer for ASR}
\label{sec:lrt}

\begin{figure}[tbp]
\centerline{\includegraphics[width=\linewidth]{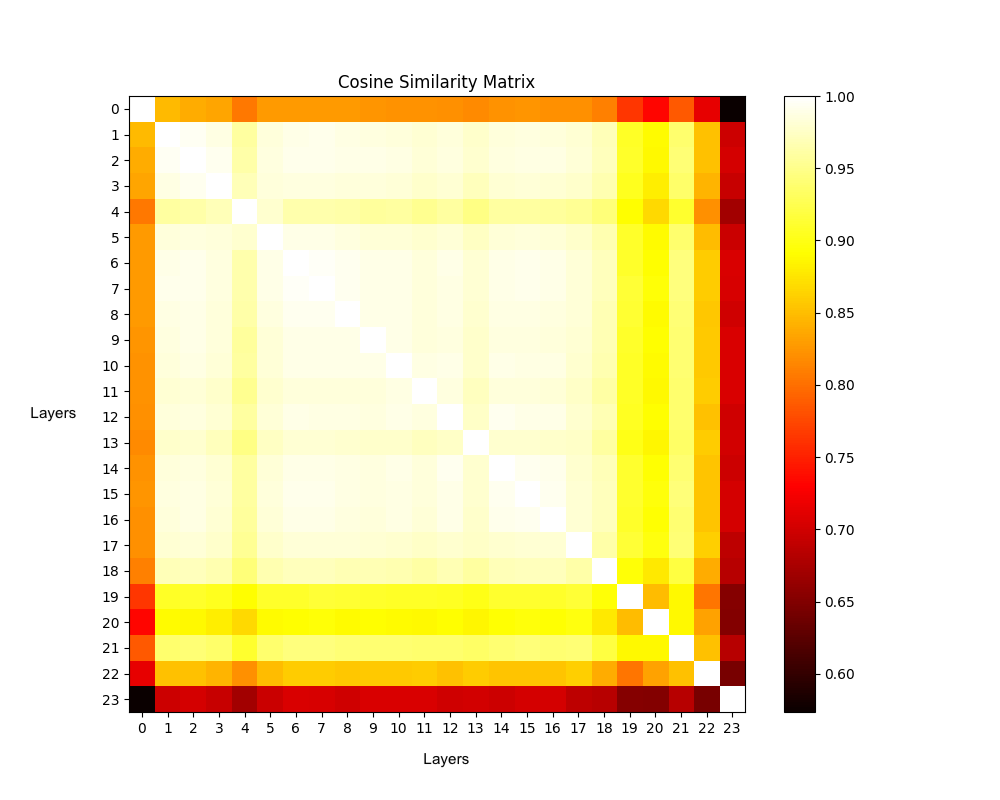}}
\caption{Cosine similarity matrix of layer-wise outputs from the Whisper-medium encoder, computed using identical input across all layers. Each element (i, j) in the matrix represents the cosine similarity between the outputs of layers i and j.
}
\label{fig:WhisperEncoder}
\end{figure}

\begin{figure*}[t]
\centerline{\includegraphics[width=0.8\linewidth]{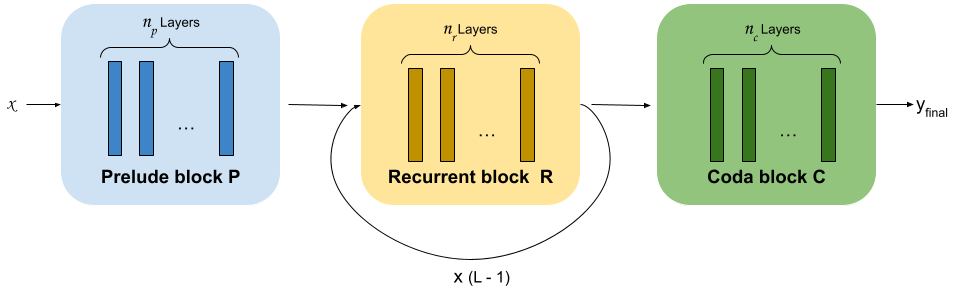}}
\caption{Diagram  of the Latent-Recursive-Transformer architecture. Each block is composed of several sub-layers. The Prelude block (P) encodes the inputs into a latent representation. The Recurrent block (R), shared across multiple steps, iteratively refines this latent representation. Finally, the Coda block (C) decodes the latent state to produce the output. 
}
\label{fig:archi}
\end{figure*}
% Whisper motivation here
This work is motivated by the observation that prior research in Recursive-Transformers for ASR selects the number of loops and the shared layers arbitrarily. We argue that such choices should be guided by a deeper understanding of the model's internal dynamics. To this end, as our work focuses on the ASR encoder, we first analyze the functional role of each layer within a pre-trained ASR encoder, namely Whisper-medium \cite{radford2023robust}, following the methodology from previous NLP research \cite{sun2024transformer}.
As shown in Figure \ref{fig:WhisperEncoder}, the Whisper-medium encoder reveals that middle layers yield highly similar representations, while edge layers produce distinct outputs. While this phenomenon has been documented in NLP decoders \cite{sun2024transformer}, to the best of our knowledge, it is reported here for the first time in the context of ASR encoders. This confirms that middle layers are largely redundant and can be compressed into a shared recurrent block.

The Latent-Recursive-Transformer is a variant of the Recursive-Transformer \cite{geiping2025scaling, reid2021subformer,yu2025mesh} that applies depth recurrence specifically within the latent space to facilitate latent reasoning during test-time computations \cite{geiping2025scaling, DBLP:conf/iclr/SaunshiDLKR25}. 
%This work is motivated by the observation that prior Recurrent-Transformers research in both NLP and ASR often selects the number of loops and shared layers arbitrarily. We argue that such choices should be guided by a deeper understanding of the model's dynamics. To this end, as our work focuses on the ASR encoder, we first analyse the role of each layer within the Whisper encoder \cite{radford2023robust}, a state-of-the-art large ASR encoder, by computing the similarity of their outputs for identical inputs, following the method from previous NLP research \cite{sun2024transformer}. As shown in Figure \ref{fig:WhisperEncoder}, the middle layers yield highly similar representations, whereas the first and final layers produce distinct outputs, a finding consistent with prior analyses of NLP decoders \cite{sun2024transformer}, but, to the best of our knowledge, reported here for the first time in ASR. This insight motivates our use of a Latent-Recursive-Transformer architecture in our experiments.

% Recursive Transformer Layer
%The Latent-Recursive-Transformer is a variant of the Recursive-Transformer \cite{geiping2025scaling,reid2021subformer}, with direct application of depth recurrence within the latent space to facilitate latent reasoning during test-time computations  \cite{geiping2025scaling,DBLP:conf/iclr/SaunshiDLKR25}. This approach has yielded improved results on reasoning and mathematical benchmarks. 
The structure of the Latent-Recursive-Transformer differs from the regular Recursive-Transformer model as it divides the model into three functional groups: the Prelude ($P$), the core Recurrent block ($R$), and the Coda ($C$), which correspond to the initial, middle, and final blocks of layers, respectively. The Prelude block encodes inputs into a latent representation; the Recurrent block, shared across multiple steps, iteratively refines this representation; and the Coda block decodes the latent state to produce the final output. This architecture, illustrated in Figure \ref{fig:archi},  can be formally expressed as follows:
\begin{align}
   & \mathbf{y_0} = P(\mathbf{x}) \\
   & \mathbf{y_{i}} = R(\mathbf{y_{i-1}})  \ \ \ \ \ \text{ for } i \in \{1,..,L\} \\
   & \mathbf{y_{final}} = C(\mathbf{y_L}).
\end{align}
Here, $ \mathbf{x}$ and $\mathbf{y_{final}}\in \mathbb{R}^{t \times d} $ represent the input and final output of the Latent-Recursive-Transformer model, where $ t $ denotes the sequence length, $d$ the embedding dimension, and  $L$ indicates the number of loops of the recurrent block. It is important to note that each of the blocks $P$, $R$, and $C$ comprises $ n_p $, $ n_r $, and $ n_c $ layers, respectively, with  $ n_p, n_r, n_c \geq 0 $. 
In particular, when $n_p = n_c = 0$, it signifies the absence of the Prelude and Coda blocks, resulting in a regular Recursive-Transformer. 
Additionally, $ L = 1 $ 
denotes the case in which the data flows through the recurrent block a single time, corresponding to the non-recursive model. 

Despite the formal flexibility of this framework, the empirical impact of the recursion depth $L$ and the specific layer allocation across functional blocks remains under-explored in the context of ASR encoders. It is currently unclear how varying the number of loops affects the model's performance. Consequently, this work systematically investigates these dynamics to identify the optimal balance between these aspects.
\section{Experimental setting}

\subsection{Corpora}
Our experiments primarily use the LibriSpeech corpus \cite{panayotov2015librispeech}, a widely adopted ASR benchmark comprising roughly 1,000 hours of English audiobook speech from LibriVox \cite{kearns2014librivox}. We follow the standard partitions with 960 hours for training and two test set, test-clean and test-other, each containing 5 hours. To evaluate the generalisability of this study across languages, we also employ AISHELL-1 \cite{bu2017aishell}, a 150-hour Mandarin corpus of clean read speech recorded in controlled indoor conditions, with 5 hours for test.

\subsection{Implementation details}
All experiments were conducted using the SpeechBrain toolkit \cite{speechbrain_v1}. To isolate the effects of recurrence on the acoustic representation, our study focuses exclusively on the encoder, with the number of Transformer layers $n_p+n_r+n_c$ ranging from 1 to 24. The decoder is consistently composed of 24 Transformer layers to maintain a stable baseline for comparison. Consequently, parameters associated with the decoder are not included in the parameter counts reported in our results.

Furthermore, while adaptive early stopping mechanisms could significantly improve inference efficiency, their inclusion would introduce a dynamic variable that could confound our analysis of how the fixed loop count $L$ impacts model performance. To ensure a clear and controlled investigation of the relationship between recursion depth and recognition accuracy, we chose to keep L constant for each experimental run. Consequently, dynamic depth strategies were not explored in the present study and are reserved for future work.

For the decoding process, a Transformer language model was employed, which had been trained on 10 million words derived from the transcriptions of the LibriSpeech dataset.

For the AISHELL-1 dataset, we used a smaller Transformer architecture with dimension $d$ of 256, using up to 12 layers for the encoder and 6 for the decoder.

The training of all configurations was conducted over 60 epochs, employing a learning rate of $8 \times 10^{-3}$ using a combination of Connectionist Temporal Classification (CTC) \cite{graves2012connectionist} and Sequence-to-Sequence (Seq2Seq) loss functions with respective weights of 0.3 and 0.7.

\section{Results}

\subsection{Recursive-Transformer configurations}
\begin{table}[t]
    \centering
    \caption{Evaluation of Recursive-Transformer (R) and Latent-Recursive-Transformer (L). Configurations in WER \% clean (other) for Librispeech. Besides L3, all configurations use the same amount of FLOPs.}
    \begin{tabular}{@{}cc|ccccccc@{}}
        \hline
         Config & $\{n_r, L, n_p, n_c\}$  & WER \% & $\lVert\text{Param}\lVert$ & $\Delta$ size \\ \hline
        B & \{24, 1, 0, 0\}  & 2.12 (4.76) & 75.6M & - \\ \hline
        R1 & \{12, 2, 0, 0\}  & 2.15 (5.03) & 37.8M & -50\% \\
        R2 & \{6, 4, 0, 0\}   & 2.25 (5.23) & 18.9M & -75\% \\
        R3 & \{4, 6, 0, 0\}   & 2.32 (5.24) & 12.6M & -83\%\\
        R4 & \{1, 24, 0, 0\}  & 2.78 (6.49) & 3.1M  & -95\%\\ \hline
        L1 & \{4, 5, 2, 2 \}   & 2.16 (4.92) & 25.2M & -66\%\\ 
        L2 & \{1, 20, 2, 2\}   & 2.31 (5.30) & 15.7M & -79\% \\ \hline
        L3 & \{20, 5, 2, 2\}   & \textbf{2.03 (4.73)} & 75.6M & - \\ \hline
    \end{tabular}
    \label{tab:model_configs}
\end{table}

We evaluate the influence of different Recursive-Transformer configurations, parameterised by the tuple $\{n_r, L, n_p, n_c\}$. %where $n_r$ denotes the number of layers in the recurrent block $R$, $L$ the recursion depth, and $n_p$ and $n_c$ the number of layers in the $P$ and $C$ blocks of the Latent-Recursive Transformer, respectively.
To ensure fair comparison, all configurations process the same number of layers, such that $n_p + L \times n_r + n_c = 24$, unless stated otherwise.

The baseline configuration (B) consists of 24 unique layers, corresponding to a single pass through the recurrent block. It achieves a word error rate (WER) of 2.12\% (4.76\%), for test-clean and test-other respectively, with 75.6M parameters.

When $n_r$ is reduced and $L$ increased (configurations R1 and R2), WER rises slightly to 2.15\% (5.03\%) and 2.25\% (5.23\%), while parameter counts drop to 37.8M and 18.9M, respectively. This indicates that model size can be reduced by more than half with only marginal performance degradation. Further reduction (R3) yields a WER of 2.32\% (5.24\%) with 12.6M parameters—an 83\% reduction relative to the baseline—while R4, comprising a single layer looped 24 times, records the highest WER of 2.78\% (6.49\%) with only 3.1M parameters (a 95\% reduction). These results suggest that extreme recurrence may compromise representational capacity, though performance remains competitive given the drastic reduction in parameters.

\subsection{Latent-Recursive-Transformer configurations}

Building on the findings in Section \ref{sec:lrt}, we evaluate the Latent-Recursive-Transformer by incorporating non-shared layers in the $P$ and $C$ blocks, motivated by the distinct representational roles of the encoder’s edge layers (Figure \ref{fig:WhisperEncoder}). Specifically, we introduce two unique layers in each block.

Modified versions of R3 and R4, denoted L1 and L2 respectively, achieve WERs of 2.16\% (4.92\%) and  2.31\% (5.30\%) with parameter counts of 25.2M and 15.7M, respectively (Table \ref{tab:model_configs}). These results demonstrate that augmenting Recursive-Transformer architectures with latent non-shared layers consistently improves recognition performance, highlighting the advantages of the Latent-Recursive design.

We further evaluated a Latent-Recursive configuration (L3, \{20,5,2,2\}) with the same number of parameters as baseline B, achieving 2.03\%  (4.73\%) WER, surpassing standard Transformers $B$ under parameter-matched settings.

\subsection{Influence of the number of recurrence loops}
\begin{table}[t]
    \centering
    \caption{Impact of the number of recurrence loops in $L1$ configuration for Librispeech in WER \%.}
    \begin{tabular}{@{}l|ccccccc@{}}
        \hline
        Config & $\{n_r, L, n_p, n_c\}$  & WER  \% clean (other)  \\ \hline
        $L1_1$ &\{4, 1, 2, 2\} & 2.44 (5.63) \\ %2.39 (5.47) \\
        $L1_2$ & \{4, 2, 2, 2\} & 2.23 (5.12) \\
        $L1$ &\{4, 5, 2, 2\} & 2.16 \textbf{(4.92)} \\
        $L1_{10}$ & \{4, 10, 2, 2\}& 2.18 (4.94) \\
        $L1_{20}$ &\{4, 20, 2, 2\}& 2.31 (5.30) \\ \hline %2.20 (5.00) \\ \hline
        $L1_{1\dots20}$ &\{4, \{1,\dots,20\}, 2, 2\} & 2.22 (5.05)  \\
        $L1_{1\rightarrow20}$ &\{4, 1$\rightarrow $20 , 2, 2\} & \textbf{2.13} (4.97)\\ \hline
    \end{tabular}
    \label{tab:Loops}
\end{table}

\begin{figure}
    \centering
    \includegraphics[width=0.8\linewidth]{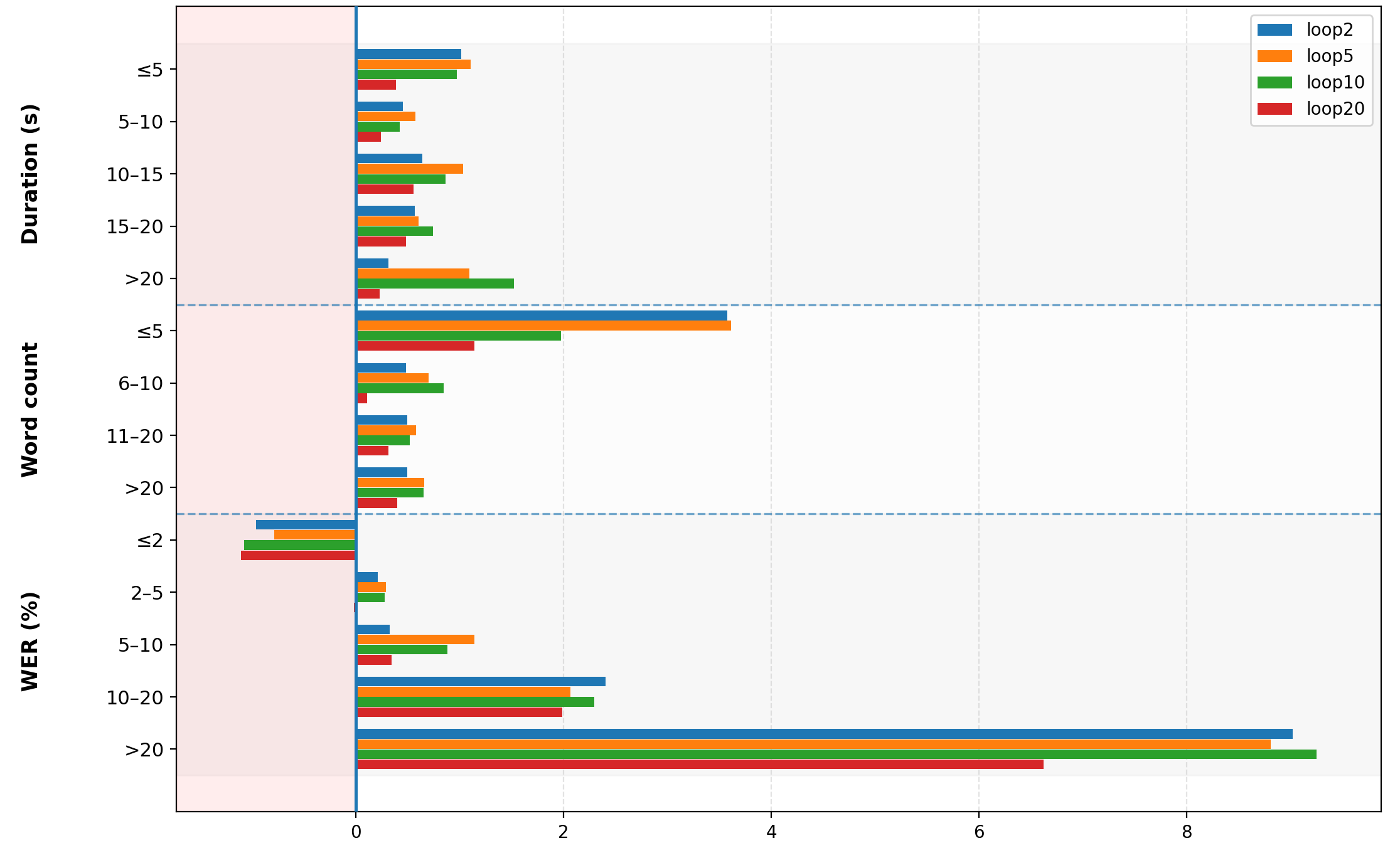}
    \caption{WER change relative to the baseline model for utterance duration, transcript length (word-count), and baseline difficulty (per-utterance baseline WER bins) for test-other. Bars show the mean improvement $\Delta\mathrm{WER}=\mathrm{WER}_{\text{L1}_1}-\mathrm{WER}_{L1_{\text{loop}}}$ for $L1_{loop}$ variants. Positive values indicate lower WER than baseline. The shaded region highlights negative improvement (degradation).}
    \label{fig:barh}
\end{figure}
We examine the impact of the number of recurrence loops $L$ in the Latent-Recursive-Transformer. Table \ref{tab:Loops} reports results obtained with varying $L$ while keeping the number of unique layers fixed. Note that the computational budget differs across configurations, as data passes through different numbers of (potentially repeated) layers. The $L1_1$ configuration, corresponding to a single pass through block $R$ (no recurrence), produced the weakest performance, with WERs of 2.44\% (test-clean) and 5.63\% (test-other), underscoring the importance of recurrence. Introducing two loops ($L1_2$) improved performance to 2.23\% and 5.12\%, while five loops ($L1$) yielded the best results: 2.16\% and 4.92\%. This highlights the role of recurrence depth in enhancing representational capacity.
However, increasing the number of loops beyond this degraded performance. For example, $L1_{10}$ and $L1_{20}$ resulted in WERs of 2.18\% (4.94\%) and 2.31\% (5.30\%), respectively. These findings suggest that excessive looping may overcomplicate training and reduce generalisation.

Figure \ref{fig:barh} shows that depth recurrence concentrates its gains on difficult utterances, with the largest average WER reductions for high-error cases ($\text{WER}_{\text{base}} > 20\%$) and long segments ($>20 s$), consistent with improved global consistency when the first-pass hypothesis is weak. In contrast, low-error bins ($\text{WER}_{\text{base}} \le 5\%$, especially $\le 2\%$) exhibit little benefit and occasional degradation, suggesting over-correction. Very short transcripts ($\le 5$ words) are also comparatively unstable under recurrence, likely because small edits produce large relative changes.%Figure \ref{fig:barh} demonstrates a clear performance-complexity trade-off: depth recurrence significantly improves "hard" utterances (baseline WER $>$ 20\%) and long-duration segments ($>$ 20s), In contrast, easy utterances (baseline WER $\leq$ 5\%) see little benefit and often slight degradation, suggesting over-correction when the initial hypothesis is already accurate. Finally, very short transcripts ($\leq$ 5 words) are consistently harmed, likely because small changes can disproportionately alter the entire hypothesis.

In summary, the choice of recursion depth is critical: too few or too many loops both impair performance. These results align with previous findings on the depth of Transformers \cite{csordas2025language}.
\subsection{Training with a Dynamic Number of Loops}
Subsequently, we evaluate using a dynamic number loop $L$ during training. Particularly, we explore two such strategies for the Latent-Recursive-Transformer. First, the Length Strategy ($L1_{1\dots20}$) assigns each utterance $n$ a loop count $L_n$ proportional to its duration, as longer utterances might need more iterative refinement, ranging from $L_{\min}=1$ to $L_{\max}=20$:

\begin{equation}
L_n = \lfloor L_{\min} + \frac{(len(n) - min\_len)(L_{\max} - L_{\min})}{max\_len - min\_len} \rfloor
\end{equation}

Then, the Schedule Strategy ($L1_{1\rightarrow20}$) gradually increases $L$ during training, starting from 1 and reaching 20, while using $L=20$ for inference:
\begin{equation}  
L = \min \left(1 + \left\lfloor \tfrac{\text{current\_epoch}}{3} \right\rfloor, 20 \right)
\end{equation}

As shown in Table \ref{tab:Loops}, the Length Strategy failed to improve performance compared to the fixed-$L$ training with 20 loops. In contrast, the Schedule Strategy achieves the best test-clean performance among all $L1$ variants.

\subsection{Generalisation across architectures and datasets}

\begin{table}[t]
    \centering
    \caption{Extended evaluation: AISHELL-1 (CER \%) and LibriSpeech Branchformer (WER \%).}
    \resizebox{\columnwidth}{!}{%
    \begin{tabular}{@{}c|c|c|c|c@{}}
        \hline
        \textbf{Config} & $\{ n_r, L, n_p, n_c\}$ & \textbf{Error rate \%} & $\lVert\text{Param}\rVert$ & $\Delta$ size \\
        \hline
        \multicolumn{5}{c}{\textbf{AISHELL-1 (CER\%) - Transformer}} \\ \hline
         B2  & $\{12, 1, 0, 0\}$ & \textbf{6.43} & 15.7M & -- \\
         L4  & $\{2, 4, 2, 2\}$  & 6.50 & 7.8M & -50\% \\
        \hline
        \multicolumn{5}{c}{\textbf{LibriSpeech (WER\%: clean (other)) --- Branchformer}} \\ \hline
         B3  & $\{24, 1, 0, 0\}$ & \textbf{1.97} (4.86) & 101.6M & -- \\
         L5  & $\{4, 5, 2, 2\}$  & 2.04 \textbf{(4.62)} & 33.6M & -66\% \\
        \hline
    \end{tabular}%
    }
    \label{tab:models_branchformer}
\end{table}
To evaluate the generalisability of this study, we extend experiments to AISHELL-1 and the Branchformer \cite{peng2022branchformer} on LibriSpeech (Table \ref{tab:models_branchformer}). On AISHELL-1, the baseline Transformer (B2) achieves 6.43\% character error rate (CER) with 15.7M parameters, while the Latent-Recursive variant (L4) maintains comparable accuracy (6.50\%) with 50\% fewer parameters. On LibriSpeech, the Branchformer baseline (B3) records 1.97\% and 4.86\% WER on test-clean and test-other, respectively, whereas the Latent-Recursive Branchformer (L5) achieves 2.04\% and 4.62\% with 66\% fewer parameters. These results demonstrate that the Latent-Recursive approach preserves competitive performance while substantially reducing model size across datasets and architecture variants.

\section{Conclusion and future work}
In this work, we investigated Recursive- and Latent-Recursive-Transformer encoders for ASR, showing that they preserve performance with substantially fewer parameters and can outperform standard Transformers under parameter-matched settings, particularly for the Latent-Recursive architecture. Our experiments highlight the critical role of recurrence depth. We further demonstrated that these architectures generalise well across Transformer variants like Branchformer and datasets. Future work will explore methods for automatically selecting the optimal number of recurrence loops $L$ and incorporating low-rank adaptations within the shared layers of the recurrent block $R$.

\section{Acknowledgements}
This work was funded/supported by Portuguese national funds through Fundação para a Ciência e a Tecnologia, I.P. (FCT) under projects UID/50021/2025 and UID/PRR/50021/2025, and by the Portuguese Recovery and Resilience Plan and NextGenerationEU European Union funds under project C644865762-00000008 (Accelerat.AI).
\bibliographystyle{IEEEtran}
\bibliography{mybib}

@article{Transformer,
  title={Attention is all you need},
  author={Vaswani, Ashish and Shazeer, Noam and Parmar, Niki and Uszkoreit, Jakob and Jones, Llion and Gomez, Aidan N and Kaiser, {\L}ukasz and Polosukhin, Illia},
  journal={Advances in neural information processing systems},
  volume={30},
  year={2017}
}

@article{gondi2021performance,
  title={Performance and efficiency evaluation of ASR inference on the edge},
  author={Gondi, Santosh and Pratap, Vineel},
  journal={Sustainability},
  volume={13},
  number={22},
  pages={12392},
  year={2021},
  publisher={MDPI}
}

@inproceedings{radford2023robust,
  title={Robust speech recognition via large-scale weak supervision},
  author={Radford, Alec and Kim, Jong Wook and Xu, Tao and Brockman, Greg and McLeavey, Christine and Sutskever, Ilya},
  booktitle={International conference on machine learning},
  pages={28492--28518},
  year={2023},
  organization={PMLR}
}

@article{chen2025owls,
  title={OWLS: Scaling laws for multilingual speech recognition and translation models},
  author={Chen, William and Tian, Jinchuan and Peng, Yifan and Yan, Brian and Yang, Chao-Han Huck and Watanabe, Shinji},
  journal={arXiv:2502.10373},
  year={2025}
}

@article{Kaplan2020ScalingLF,
  title={Scaling Laws for Neural Language Models},
  author={Jared Kaplan and Sam McCandlish and T. J. Henighan and Tom B. Brown and Benjamin Chess and Rewon Child and Scott Gray and Alec Radford and Jeff Wu and Dario Amodei},
  journal={arXiv:2001.08361},
  year={2020},
  volume={abs/2001.08361},
  url={https://api.semanticscholar.org/CorpusID:210861095}
}

@inproceedings{pham2018efficient,
  title={Efficient neural architecture search via parameters sharing},
  author={Pham, Hieu and Guan, Melody and Zoph, Barret and Le, Quoc and Dean, Jeff},
  booktitle={International conference on machine learning},
  pages={4095--4104},
  year={2018},
  organization={PMLR}
}

@inproceedings{xue2022go,
  title={Go wider instead of deeper},
  author={Xue, Fuzhao and Shi, Ziji and Wei, Futao and Lou, Yuxuan and Liu, Yong and You, Yang},
  booktitle={Proceedings of the AAAI Conference on Artificial Intelligence},
  volume={36},
  number={8},
  pages={8779--8787},
  year={2022}
}

@inproceedings{pires2023one,
    title = "One Wide Feedforward Is All You Need",
    author = "Pires, Telmo  and
      Vilarinho Lopes, Ant{\'o}nio  and
      Assogba, Yannick  and
      Setiawan, Hendra",
    editor = "Koehn, Philipp  and
      Haddow, Barry  and
      Kocmi, Tom  and
      Monz, Christof",
    booktitle = "Proceedings of the Eighth Conference on Machine Translation",
    month = dec,
    year = "2023",
    address = "Singapore",
    publisher = "Association for Computational Linguistics",
    url = "https://aclanthology.org/2023.wmt-1.98/",
    doi = "10.18653/v1/2023.wmt-1.98",
    pages = "1031--1044"
}

@inproceedings{UT2019,title	= {Universal Transformers},author	= {Mostafa Dehghani and Stephan Gouws and Oriol Vinyals and Jakob Uszkoreit and Lukasz Kaiser},year	= {2019},URL	= {https://openreview.net/pdf?id=HyzdRiR9Y7},booktitle={ICLR 2019}, 
}

@article{graves2016adaptive,
  title={Adaptive computation time for recurrent neural networks},
  author={Graves, Alex},
  journal={arXiv preprint arXiv:1603.08983},
  year={2016}
}

@inproceedings{bae2024relaxed,
    title={Relaxed Recursive Transformers: Effective Parameter Sharing with Layer-wise LoRA},
    author={Sangmin Bae and Adam Fisch and Hrayr Harutyunyan and Ziwei Ji and Seungyeon Kim and Tal Schuster},
    booktitle={International Conference on Learning Representations},
    year={2025}
}

@article{sun2024transformer,
  title={Transformer Layers as Painters},
  author={Sun, Qi and Pickett, Marc and Nain, Aakash Kumar and Jones, Llion},
  journal={arXiv:2407.09298},
  year={2024}
}

@article{reid2021subformer,
  title={Subformer: Exploring weight sharing for parameter efficiency in generative transformers},
  author={Reid, Machel and Marrese-Taylor, Edison and Matsuo, Yutaka},
  journal={arXiv:2101.00234},
  year={2021}
}

@inproceedings{
geiping2025scaling,
title={Scaling up Test-Time Compute with Latent Reasoning: A Recurrent Depth Approach},
author={Jonas Geiping and Sean Michael McLeish and Neel Jain and John Kirchenbauer and Siddharth Singh and Brian R. Bartoldson and Bhavya Kailkhura and Abhinav Bhatele and Tom Goldstein},
booktitle={ES-FoMo III: 3rd Workshop on Efficient Systems for Foundation Models},
year={2025},
url={https://openreview.net/forum?id=D6o6Bwtq7h}
}

@article{lan2019albert,
  title={Albert: A lite bert for self-supervised learning of language representations},
  author={Lan, Zhenzhong and Chen, Mingda and Goodman, Sebastian and Gimpel, Kevin and Sharma, Piyush and Soricut, Radu},
  journal={arXiv:1909.11942},
  year={2019}
}

@inproceedings{DBLP:conf/iclr/SaunshiDLKR25,
  author       = {Nikunj Saunshi and
                  Nishanth Dikkala and
                  Zhiyuan Li and
                  Sanjiv Kumar and
                  Sashank J. Reddi},
  title        = {Reasoning with Latent Thoughts: On the Power of Looped Transformers},
  booktitle    = {The Thirteenth International Conference on Learning Representations,
                  {ICLR} 2025, Singapore, April 24-28, 2025},
  publisher    = {OpenReview.net},
  year         = {2025},
  url          = {https://openreview.net/forum?id=din0lGfZFd},
  bibsource    = {dblp computer science bibliography, https://dblp.org}
}

@inproceedings{wang2024residualtransformer,
author = {Wang, Yiming and Li, Jinyu},
title = {ResidualTransformer: Residual Low-Rank Learning with Weight-Sharing for Transformer Layers},
organization = {ICASSP},
year = {2024},
month = {April},
publisher = {IEEE},
}

@inproceedings{
fan2025looped,
title={Looped Transformers for Length Generalization},
author={Ying Fan and Yilun Du and Kannan Ramchandran and Kangwook Lee},
booktitle={ICLR 2025},
year={2025},
url={https://openreview.net/forum?id=2edigk8yoU}
}

@inproceedings{shim24_interspeech,
  title     = {Leveraging Adapter for Parameter-Efficient ASR Encoder},
  author    = {Kyuhong Shim and Jinkyu Lee and Hyunjae Kim},
  year      = {2024},
  booktitle = {Interspeech 2024},
  pages     = {2380--2384},
  doi       = {10.21437/Interspeech.2024-334},
  issn      = {2958-1796},
}

@inproceedings{chi2021audio,
  title={Audio albert: A lite bert for self-supervised learning of audio representation},
  author={Chi, Po-Han and Chung, Pei-Hung and Wu, Tsung-Han and Hsieh, Chun-Cheng and Chen, Yen-Hao and Li, Shang-Wen and Lee, Hung-yi},
  booktitle={2021 IEEE Spoken Language Technology Workshop (SLT)},
  pages={344--350},
  year={2021},
  organization={IEEE}
}

@article{wei2023sim,
  title={Sim-T: Simplify the transformer network by multiplexing technique for speech recognition},
  author={Wei, Guangyong and Duan, Zhikui and Li, Shiren and Yang, Guangguang and Yu, Xinmei and Li, Junhua},
  journal={arXiv:2304.04991},
  year={2023}
}

@inproceedings{panayotov2015librispeech,
  title={Librispeech: an asr corpus based on public domain audio books},
  author={Panayotov, Vassil and Chen, Guoguo and Povey, Daniel and Khudanpur, Sanjeev},
  booktitle={ICASSP},
  pages={5206--5210},
  year={2015},
  organization={IEEE}
}

@article{speechbrain_v1,
  author  = {Mirco Ravanelli and Titouan Parcollet and Adel Moumen and Sylvain de Langen and Cem Subakan and Peter Plantinga and Yingzhi Wang and Pooneh Mousavi and Luca Della Libera and Artem Ploujnikov and Francesco Paissan and Davide Borra and Salah Zaiem and Zeyu Zhao and Shucong Zhang and Georgios Karakasidis and Sung-Lin Yeh and Pierre Champion and Aku Rouhe and Rudolf Braun and Florian Mai and Juan Zuluaga-Gomez and Seyed Mahed Mousavi and Andreas Nautsch and Ha Nguyen and Xuechen Liu and Sangeet Sagar and Jarod Duret and Salima Mdhaffar and Ga{{\"e}}lle Laperri{{\`e}}re and Mickael Rouvier and Renato De Mori and Yannick Est{{\`e}}ve},
  title   = {Open-Source Conversational AI with SpeechBrain 1.0},
  journal = {Journal of Machine Learning Research},
  year    = {2024},
  volume  = {25},
  number  = {333},
  url     = {http://jmlr.org/papers/v25/24-0991.html}
}

@article{csordas2025language,
  title={Do Language Models Use Their Depth Efficiently?},
  author={Csord{\'a}s, R{\'o}bert and Manning, Christopher D and Potts, Christopher},
  journal={arXiv:2505.13898},
  year={2025}
}

@inproceedings{peng2022branchformer,
  title={Branchformer: Parallel mlp-attention architectures to capture local and global context for speech recognition and understanding},
  author={Peng, Yifan and Dalmia, Siddharth and Lane, Ian and Watanabe, Shinji},
  booktitle={International Conference on Machine Learning},
  pages={17627--17643},
  year={2022},
  organization={PMLR}
}

@article{gholami2024ai,
  title={Ai and memory wall},
  author={Gholami, Amir and Yao, Zhewei and Kim, Sehoon and Hooper, Coleman and Mahoney, Michael W and Keutzer, Kurt},
  journal={IEEE Micro},
  volume={44},
  number={3},
  pages={33--39},
  year={2024},
  publisher={IEEE}
}

@inproceedings{bu2017aishell,
  title={Aishell-1: An open-source mandarin speech corpus and a speech recognition baseline},
  author={Bu, Hui and Du, Jiayu and Na, Xingyu and Wu, Bengu and Zheng, Hao},
  booktitle={2017 20th conference of the oriental chapter of the international coordinating committee on speech databases and speech I/O systems and assessment (O-COCOSDA)},
  pages={1--5},
  year={2017},
  organization={IEEE}
}

@inproceedings{rolland2025exploring,
  title={Exploring Shared-Weight Mechanisms in Transformer and Conformer Architectures for Automatic Speech Recognition},
  author={Rolland, Thomas and Abad, Alberto},
  booktitle={Proc. Interspeech 2025},
  pages={2885--2889},
  year={2025}
}

@article{cheng2017survey,
  title={A survey of model compression and acceleration for deep neural networks},
  author={Cheng, Yu and Wang, Duo and Zhou, Pan and Zhang, Tao},
  journal={arXiv:1710.09282},
  year={2017}
}

@article{koluguri2025granary,
  title={Granary: Speech Recognition and Translation Dataset in 25 European Languages},
  author={Koluguri, Nithin Rao and Sekoyan, Monica and Zelenfroynd, George and Meister, Sasha and Ding, Shuoyang and Kostandian, Sofia and Huang, He and Karpov, Nikolay and Balam, Jagadeesh and Lavrukhin, Vitaly and others},
  journal={arXiv:2505.13404},
  year={2025}
}

@inproceedings{alizadeh2024llm,
  title={Llm in a flash: Efficient large language model inference with limited memory},
  author={Alizadeh, Keivan and Mirzadeh, Seyed Iman and Belenko, Dmitry and Khatamifard, S and Cho, Minsik and Del Mundo, Carlo C and Rastegari, Mohammad and Farajtabar, Mehrdad},
  booktitle={Proceedings of the 62nd Annual Meeting of the Association for Computational Linguistics (Volume 1: Long Papers)},
  pages={12562--12584},
  year={2024}
}

@article{zhu2025scaling,
  title={Scaling latent reasoning via looped language models},
  author={Zhu, Rui-Jie and Wang, Zixuan and Hua, Kai and Zhang, Tianyu and Li, Ziniu and Que, Haoran and Wei, Boyi and Wen, Zixin and Yin, Fan and Xing, He and others},
  journal={arXiv preprint arXiv:2510.25741},
  year={2025}
}

@inproceedings{saunshi2025reasoning,
 author = {Saunshi, Nikunj and Dikkala, Nishanth and Li, Zhiyuan and Kumar, Sanjiv and J. Reddi, Sashank},
 booktitle = {International Conference on Learning Representations},
 editor = {Y. Yue and A. Garg and N. Peng and F. Sha and R. Yu},
 pages = {14855--14881},
 title = {Reasoning with Latent Thoughts: On the Power of Looped Transformers},
 url = {https://proceedings.iclr.cc/paper_files/paper/2025/file/2676109d49d1eb26d6bc584a8f556305-Paper-Conference.pdf},
 volume = {2025},
 year = {2025}
}

@article{bae2025mixture,
  title={Mixture-of-recursions: Learning dynamic recursive depths for adaptive token-level computation},
  author={Bae, Sangmin and Kim, Yujin and Bayat, Reza and Kim, Sungnyun and Ha, Jiyoun and Schuster, Tal and Fisch, Adam and Harutyunyan, Hrayr and Ji, Ziwei and Courville, Aaron and others},
  journal={arXiv preprint arXiv:2507.10524},
  year={2025}
}

@misc{kearns2014librivox,
  title={Librivox: Free public domain audiobooks},
  author={Kearns, Jodi},
  year={2014},
  publisher={Emerald group publishing limited}
}

@incollection{graves2012connectionist,
  title={Connectionist temporal classification},
  author={Graves, Alex},
  booktitle={Supervised sequence labelling with recurrent neural networks},
  pages={61--93},
  year={2012},
  publisher={Springer}
}

@article{yu2026spiralformer,
  title={SpiralFormer: Looped Transformers Can Learn Hierarchical Dependencies via Multi-Resolution Recursion},
  author={Yu, Chengting and Shu, Xiaobo and Wang, Yadao and Zhang, Yizhen and Wu, Haoyi and Wu, You and Long, Rujiao and Chen, Ziheng and Xu, Yuchi and Su, Wenbo and others},
  journal={arXiv preprint arXiv:2602.11698},
  year={2026}
}

@article{xu2026looping,
  title={Looping Back to Move Forward: Recursive Transformers for Efficient and Flexible Large Multimodal Models},
  author={Xu, Ruihan and Gao, Yuting and Wang, Lan and Li, Jianing and Chen, Weihao and Guo, Qingpei and Yang, Ming and Zhang, Shiliang},
  journal={arXiv preprint arXiv:2602.09080},
  year={2026}
}

@InProceedings{10.1007/978-3-031-20053-3_42,
author="Shen, Zhiqiang
and Liu, Zechun
and Xing, Eric",
editor="Avidan, Shai
and Brostow, Gabriel
and Ciss{\'e}, Moustapha
and Farinella, Giovanni Maria
and Hassner, Tal",
title="Sliced Recursive Transformer",
booktitle="Computer Vision -- ECCV 2022",
year="2022",
publisher="Springer Nature Switzerland",
address="Cham",
pages="727--744",
isbn="978-3-031-20053-3"
}

@inproceedings{tang2024beyond,
  title={Beyond Universal Transformer: block reusing with adaptor in Transformer for automatic speech recognition},
  author={Tang, Haoyu and Liu, Zhaoyi and Zeng, Chang and Li, Xinfeng},
  booktitle={International Symposium on Neural Networks},
  pages={69--79},
  year={2024},
  organization={Springer}
}

@article{yu2025mesh,
  title={MeSH: Memory-as-State-Highways for Recursive Transformers},
  author={Yu, Chengting and Shu, Xiaobo and Wang, Yadao and Zhang, Yizhen and Wu, Haoyi and Li, Jiaang and Long, Rujiao and Chen, Ziheng and Xu, Yuchi and Su, Wenbo and others},
  journal={arXiv preprint arXiv:2510.07739},
  year={2025}
}

@inproceedings{li25_interspeech,
  title     = {{Towards One-bit ASR: Extremely Low-bit Conformer Quantization Using Co-training and Stochastic Precision}},
  author    = {Zhaoqing Li and Haoning Xu and Zengrui Jin and Lingwei Meng and Tianzi Wang and Huimeng Wang and Youjun Chen and Mingyu Cui and Shujie Hu and Xunying Liu},
  year      = {2025},
  booktitle = {{Interspeech 2025}},
  pages     = {1973--1977},
  doi       = {10.21437/Interspeech.2025-18},
  issn      = {2958-1796},
}

@inproceedings{someki25_interspeech,
  title     = {{Context-Driven Dynamic Pruning for Large Speech Foundation Models}},
  author    = {Masao Someki and Shikhar Bharadwaj and Atharva Anand Joshi and Chyi-Jiunn Lin and Jinchuan Tian and Jee-weon Jung and Markus Müller and Nathan Susanj and Jing Liu and Shinji Watanabe},
  year      = {2025},
  booktitle = {{Interspeech 2025}},
  pages     = {1993--1997},
  doi       = {10.21437/Interspeech.2025-1193},
  issn      = {2958-1796},
}

@inproceedings{novitasari25b_interspeech,
  title     = {{Improving End-to-end Mixed-case ASR with Knowledge Distillation and Integration of Voice Activity Cues }},
  author    = {Sashi Novitasari and Takashi Fukuda and Gakuto Kurata},
  year      = {2025},
  booktitle = {{Interspeech 2025}},
  pages     = {3334--3338},
  doi       = {10.21437/Interspeech.2025-330},
  issn      = {2958-1796},
}

@inproceedings{rolland24b_interspeech,
  title     = {{Shared-Adapters: A Novel Transformer-based Parameter Efficient Transfer Learning Approach For Children’s Automatic Speech Recognition}},
  author    = {Thomas Rolland and Alberto Abad},
  year      = {2024},
  booktitle = {{Interspeech 2024}},
  pages     = {2370--2374},
  doi       = {10.21437/Interspeech.2024-1105},
  issn      = {2958-1796},
}

@inproceedings{choi2024impact,
  title={Impact of joint heat and memory constraints of mobile device in edge-assisted on-device artificial intelligence},
  author={Choi, Pyeongjun and Kim, Jeongsoo and Kwak, Jeongho},
  booktitle={Proceedings of the 2nd International Workshop on Networked AI Systems},
  pages={31--36},
  year={2024}
}

\end{document}